\documentclass[final]{IEEEtran}
\usepackage{algorithm}
\usepackage{color}
\usepackage{algpseudocode}
\usepackage{graphics}
\usepackage{epsfig}
\usepackage{cite}
\usepackage{threeparttable}
\usepackage{graphicx}
\usepackage{picinpar}
\usepackage{subfigure}
\usepackage{amssymb}
\usepackage[cmex10]{amsmath}
\usepackage{url}

\usepackage{bm}
\usepackage{setspace}

\input epsf

\begin{document}

\title{\huge{Terahertz-Band Near-Space Communications: From a Physical-Layer Perspective} }

\author{Tianqi Mao, \IEEEmembership{Member,~IEEE}, Leyi Zhang, \IEEEmembership{Student Member,~IEEE}, Zhenyu Xiao, \IEEEmembership{Senior Member,~IEEE}, Zhu Han, \IEEEmembership{Fellow,~IEEE}, and Xiang-Gen Xia, \IEEEmembership{Fellow,~IEEE}
\thanks{Tianqi Mao, Leyi Zhang and Zhenyu Xiao are with Beihang University; Zhu Han is with University of Houston; Xiang-Gen Xia is with University of Delaware; The corresponding author is Zhenyu Xiao.} %
\vspace{-3mm}} %

\maketitle
\begin{abstract}
Facilitated by rapid technological development of the near-space platform stations (NSPS), near-space communication (NS-COM) is envisioned to play a pivotal role in the space-air-ground integrated network for sixth-generation (6G) communications and beyond. In NS-COM, ultra-broadband wireless connectivity between NSPSs and various airborne/spaceborne platforms is required for a plethora of bandwidth-consuming applications, such as NSPS-based Ad hoc networking, in-flight Internet and relaying technology. However, such requirement seems to contradict with the scarcity of spectrum resources at conventional microwave frequencies, which motivates the exploitation of terahertz (THz) band ranging from $0.1$ to $10$ THz. Due to huge available bandwidth, the THz signals are capable of supporting ultra-high-rate data transmission for NS-COM over 100 Gb/s, which are naturally suitable for the near-space environment with marginal path loss. To this end, this article provides an extensive investigation on the THz-band NS-COM (THz-NS-COM) from a physical-layer perspective. Firstly, we summarize the potential applications of THz communications in the near-space environment, where the corresponding technical barriers are analyzed. Then the channel characteristics of THz-NS-COM and the corresponding modeling strategies are discussed, respectively. Afterwards, three essential research directions are investigated to surpass the technical barriers of THz-NS-COM, i.e., robust beamforming for ultra-massive antenna array, signal processing algorithms against hybrid distortions, and integrated sensing and communications. Several open problems are also provided to unleash the full potential of THz-NS-COM.

\end{abstract}
\section{Introduction}
\begin{figure*}[t!]
	\begin{center}
		\includegraphics[width=0.8\linewidth, keepaspectratio]{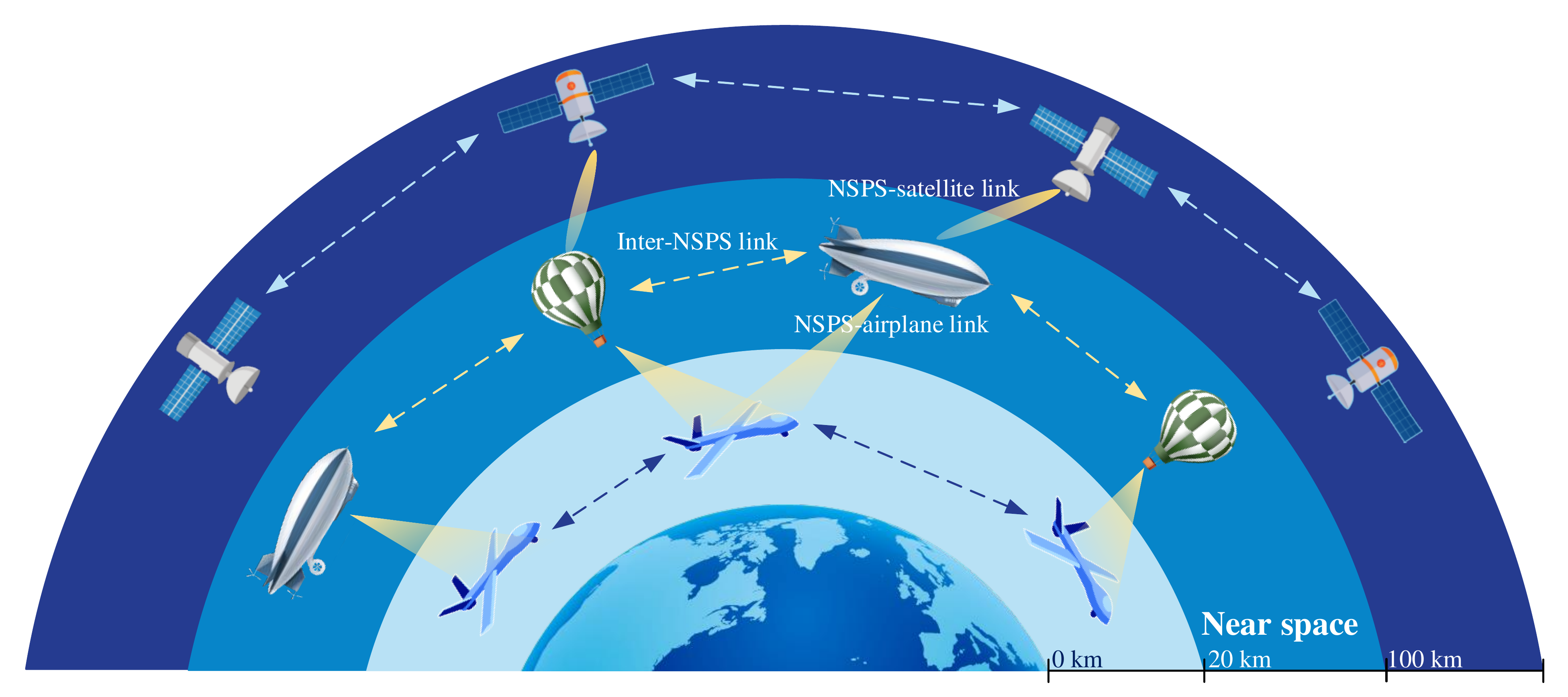}
	\end{center}
	\caption{Illustrations of THz-NS-COM networks.}
	\label{fig1}
\end{figure*}

The near space, ranging from $20$ to $100$ km in altitude, is considered as the last piece of puzzle in the sixth-generation (6G) integrated space-air-ground network architecture \cite{Zhang_conf_07}. In the recent years, the exploitation of near space has been greatly accelerated by the technical breakthrough of near-space platform station (NSPS) consisting of near-space airships and balloons. The progress includes X-Station by StratXX \cite{StationX_2022}, Loon Project by Google \cite{Loon_Google_2022}, etc. Explicitly, StratXX managed to realize mature airship products at $21$ km above the ground with super-light and super-strength materials, which enables high-speed data services with broad coverage of $10^6$ $\text{km}^2$. On the other hand, in the Loon Project, thousands of commercial near-space balloons equipped with advanced transceiving devices were deployed into the stratosphere for over 300 days' flight, constituting an Ad hoc network to ensure coverage of reliable Internet access to rural/disaster regions. Thanks to these advancements, the near-space communication (NS-COM) network can be established to provide wireless connectivity between NSPS constellations and airborne/spaceborne platforms, as illustrated in Fig. \ref{fig1}.

\begin{table*}[t!]
	\caption{Performance Comparison of Different Frequency Bands for NS-COM Applications.}	
	\centering
	\resizebox{0.8\linewidth}{!}{
		\begin{tabular}{|c|c|c|c|c|}
			\hline
			Issue & Sub-6 GHz & MmWave & Free-space optical& THz \\\hline
			Beam pattern & Broadcasting & Directional & Highly directional & Highly directional \\\hline
			Power consumption & High & Medium & Low & Low \\\hline
			Throughput & Low & Medium & High & High\\\hline
			Transmission distance & Short & Medium & Long & Long \\\hline
			Payload Size & Large & Medium & Large & Compact\\\hline
			Sensitivity to atmospheric effects & Insensitive & Moderate & Very fragile & Fragile\\\hline	
	\end{tabular}}
	\label{t1}
\end{table*}

For future 6G applications, NS-COM is expected to provide broadband Internet services to billions of terrestrial and aerial users with ubiquitous coverage \cite{Kurt_survey_21}, which necessitates ultra-fast and ultra-reliable data transmission over extremely long distance, namely, over $100$ km for inter-NSPS and NSPS-satellite links. Such requirements cannot be satisfied with traditional microwave frequencies due to scarcity of available spectrum resources and limited antenna gains. Hopefully, the terahertz (THz) band ($0.1$-$10$ THz) has emerged as a promising choice for 6G networks and beyond \cite{Chen_cm_21}, which could readily attain data transmission over $100$ Gb/s with ultra-broad available bandwidth. Due to sparsity of the atmosphere in the stratosphere and above, THz signals experience much less molecular absorption loss in the near space than its terrestrial counterpart, allowing extremely long transmission distance. Furthermore, thanks to the ultra-short wavelength of THz signals, ultra-massive antenna array is enabled to provide sufficient beamforming gain with compact size, which could support long-range directional transmission of NSPS under room limitations for communication payloads. Aside from communications, the THz spectrum also presents its superiority in sensing applications, where high resolution of range and velocity estimation can be reached with ultra-broad signal bandwidth and THz-scale center frequency \cite{Mao_tcom_22}. This realizes a number of integrated sensing and communication (ISAC) applications of NSPS, e.g., positioning/monitoring and Internet services for civil aircraft, or reconnaissance for military purposes. Table \ref{t1} compares NS-COM systems at different frequencies including sub-6 GHz, millimeter wave (mmWave), free-space optical and THz frequencies. It can be seen that, the THz spectrum is superior to the other classical counterparts from different aspects such as power consumption, throughput, transmission distance and payload size. To conclude, the unprecedented THz-band NS-COM (THz-NS-COM) network will definitely open up a plethora of bandwidth-consuming data services with ultra-broad coverage and sensing applications with superior accuracy in the future.

Despite its attractive merits, practical implementation of THz-NS-COM are facing a plethora of new technical challenges. Explicitly, the channel modeling for NS-COM data links can be sophisticated due to inconsistent property of the near-space atmosphere at different altitudes, which becomes even harder by involving the distance- and frequency-dependent characteristics of the THz long-range propagation channel; Secondly, the ``razor-sharp'' beams of THz-NS-COM systems pose great difficulty to antenna alignment, leading to degradation of the array gains, especially when communicating with airplanes/satellites of high mobility; Thirdly, THz-NS-COM systems suffer from severe hardware impairment at NSPS front-ends, which couples with the time-frequency doubly-selective channel fading, constituting deleterious nonlinear distortions on the THz signals. Finally, despite the promising prospects of ISAC, the use of ultra-broad THz spectrum causes serious expenditure issue of transceiver devices. Besides, the extremely high center frequency at THz scale exacerbates the Doppler shift effects, causing performance degradation of target sensing. At present, there have just been preliminary research efforts on the THz-band NS-COM (THz-NS-COM) framework \cite{Saeed_jsac_21,Liao_jsac_21}. Explicitly, the channel modeling issue was investigated for inter-HAPS links at the altitude of $16$ km, where desirable frequency subbands were determined according to distance-adaptive frequency selectivity of the wide-band channel and colored molecular absorption noise \cite{Saeed_jsac_21}. Additionally, \cite{Liao_jsac_21} proposed a sophisticated channel estimation and tracking design for THz-band NSPS-to-airplane links, under triple delay-beam-Doppler squint effects originated from the inherent property of THz waves and antennas.

In this paper, the great potentials and technical barriers of THz-NS-COM are extensively investigated, and some useful guidelines are provided to overcome the challenges yet to be conquered. Firstly, the characteristics of THz channels involving the near-space propagation environment are revealed, based on which different strategies of channel modeling are evaluated for THz-NS-COM. Afterwards, we discuss about several potential technologies for the settlement of aforementioned technical challenges in THz-NS-COM. Explicitly, the beamforming issue is explored under mobility of the transceiving platforms and the squint effects induced by THz-band antenna arrays. Besides, the possible solutions to the hybrid distortions caused by hardware imperfections and doubly-dispersive channel are discussed from the signal processing perspective. We also investigate the ISAC waveform design of THz-NS-COM under considerations of hardware cost and robustness to strong Doppler shifts. Correspondingly, several future research directions are also envisioned that facilitate the practical implementation of THz-NS-COM.




\begin{figure}[t!]
	\begin{center}
		\includegraphics[width=1\linewidth, keepaspectratio]{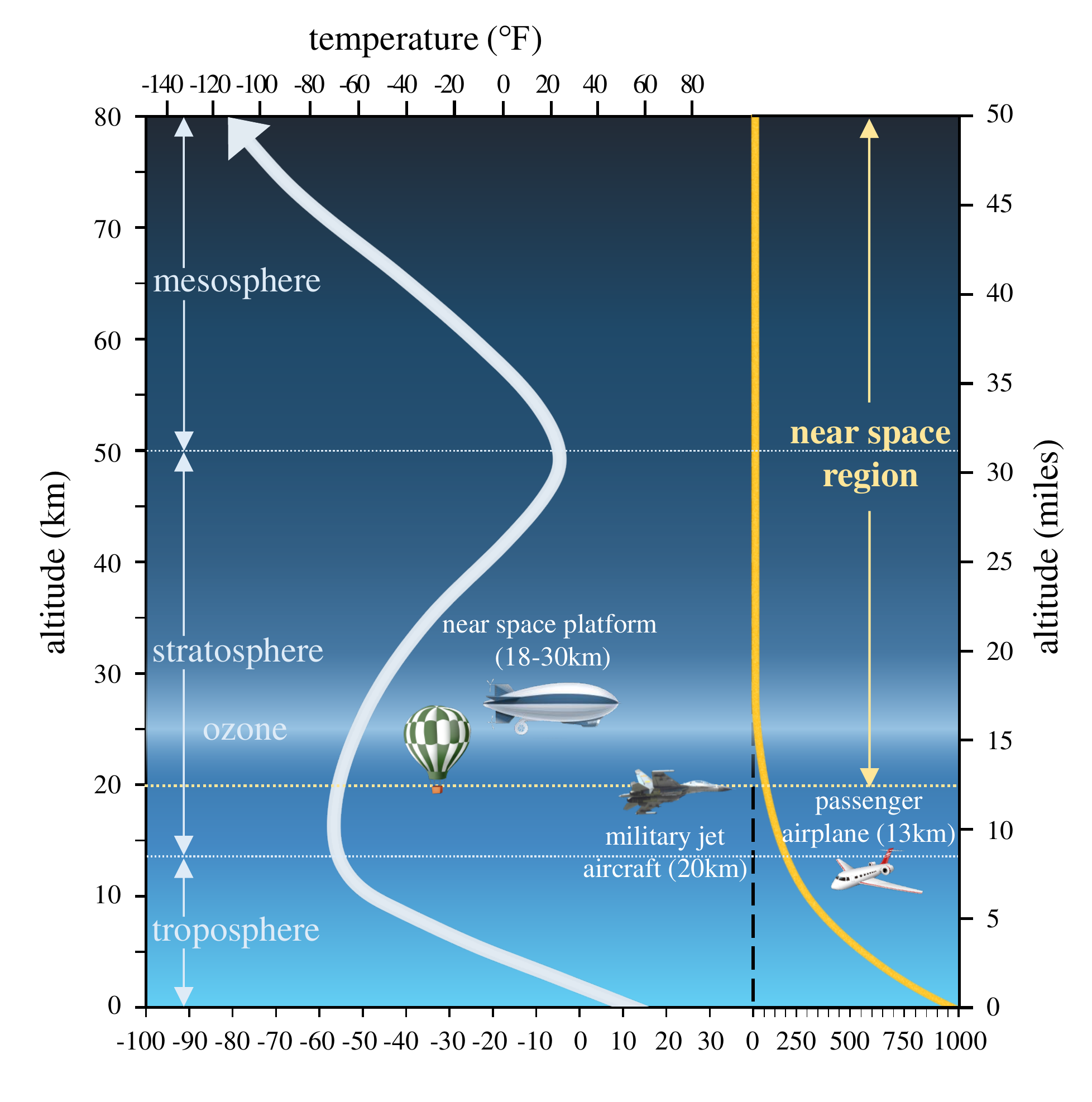}
	\end{center}
	\caption{Illustrations of the near-space communication environments \cite{britannica_2022}.}
	\label{fig2}
\end{figure} 

\section{Propagation Channel of THz-NS-COM}
Attributed to the uniqueness of near-space propagation environment, classical channel models for space/air/ground networks cannot be directly applied to explain the near-space THz channel. To this end, the characteristics of THz propagation channels in the near space, along with the mathematical modeling strategies, will be discussed below.

{\bf Propagation Environment of THz-NS-COM}: As shown in Fig. \ref{fig2}, the near-space region includes most of the stratosphere, the mesosphere and a fraction of the ionosphere, where the constituent, temperature and pressure of the atmosphere can be drastically changed with respect to the altitudes. More specifically, the ozone gas, with superior capability of ultra-violet radiation absorption, is generally distributed in the lower stratosphere. This phenomenon further causes non-monotonic changing pattern of the atmospheric temperature at the bottom of stratosphere, which declines to almost $-60$ degrees centigrade at the altitude of $20$ km due to the absorption effects of ozone layer, whilst approaches zero degree at $50$ km above the ground. Besides, despite the extremely low atmospheric pressure in most of the near space, the variations nearby the bottom of stratosphere cannot be neglected for channel modeling of the NSPS-airplane data links. To conclude, the propagation environment of THz-NS-COM presents inhomogeneous peculiarities at different altitudes, which makes the channel modeling for THz-NS-COM quite different from its terrestrial counterparts.

{\bf Channel Characteristics of THz-NS-COM}: In the terrestrial propagation environment, THz signals suffer from much more severe spreading loss and molecular absorption loss than the lower-frequency counterparts. For THz-NS-COM, on one hand, the spreading loss issue also poses great difficulty to long-range transmission approaching hundreds of km, which necessitates sufficiently high antenna gains at the transceiver. Hopefully, this requirement can be realized by establishing highly directional beams with ultra-massive THz antenna array. Nevertheless, the utilization of ``razor-sharp'' beams brings more stringent requirement on the antenna alignment accuracy at the same time. This leads to non-negligible antenna misalignment fading even for slight rotation or flutter of the NSPS, not to mention the airplane/satellite nodes with higher mobility. Furthermore, the inhomogeneous propagation environment of THz-NS-COM results in changing refractivity with respect to the altitude, which causes undesirable refraction during THz signal propagation, adding more difficulty in transceiving beam alignment.

\begin{figure*}[t!]
	\centering
	\subfigure[Delay-beam squint effects.]{
		\label{fig3a} 
		\includegraphics[width=0.45\linewidth]{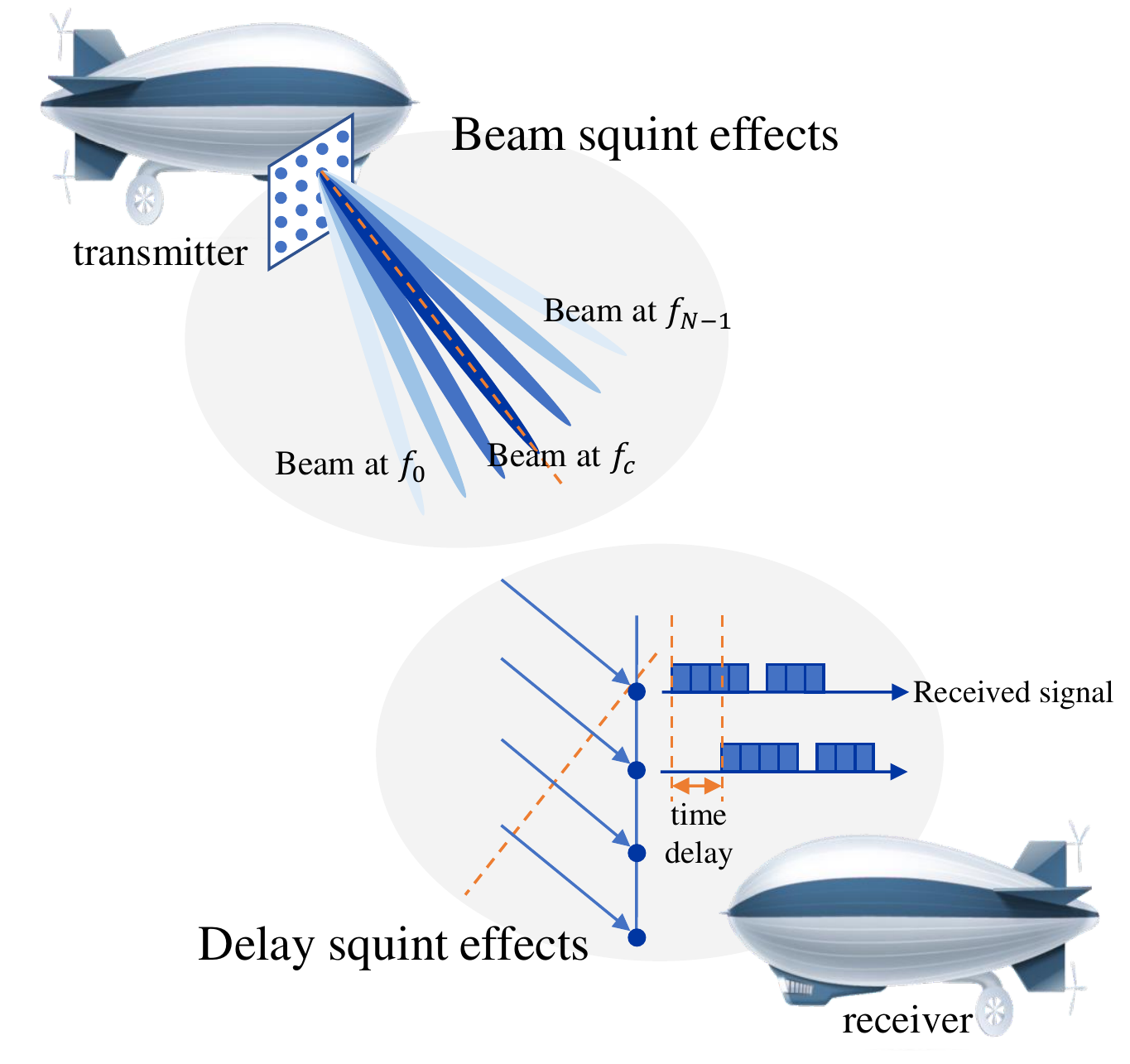}}
	\subfigure[True-time-delay network.]{
		\label{fig3b} 
		\includegraphics[width=0.4\linewidth]{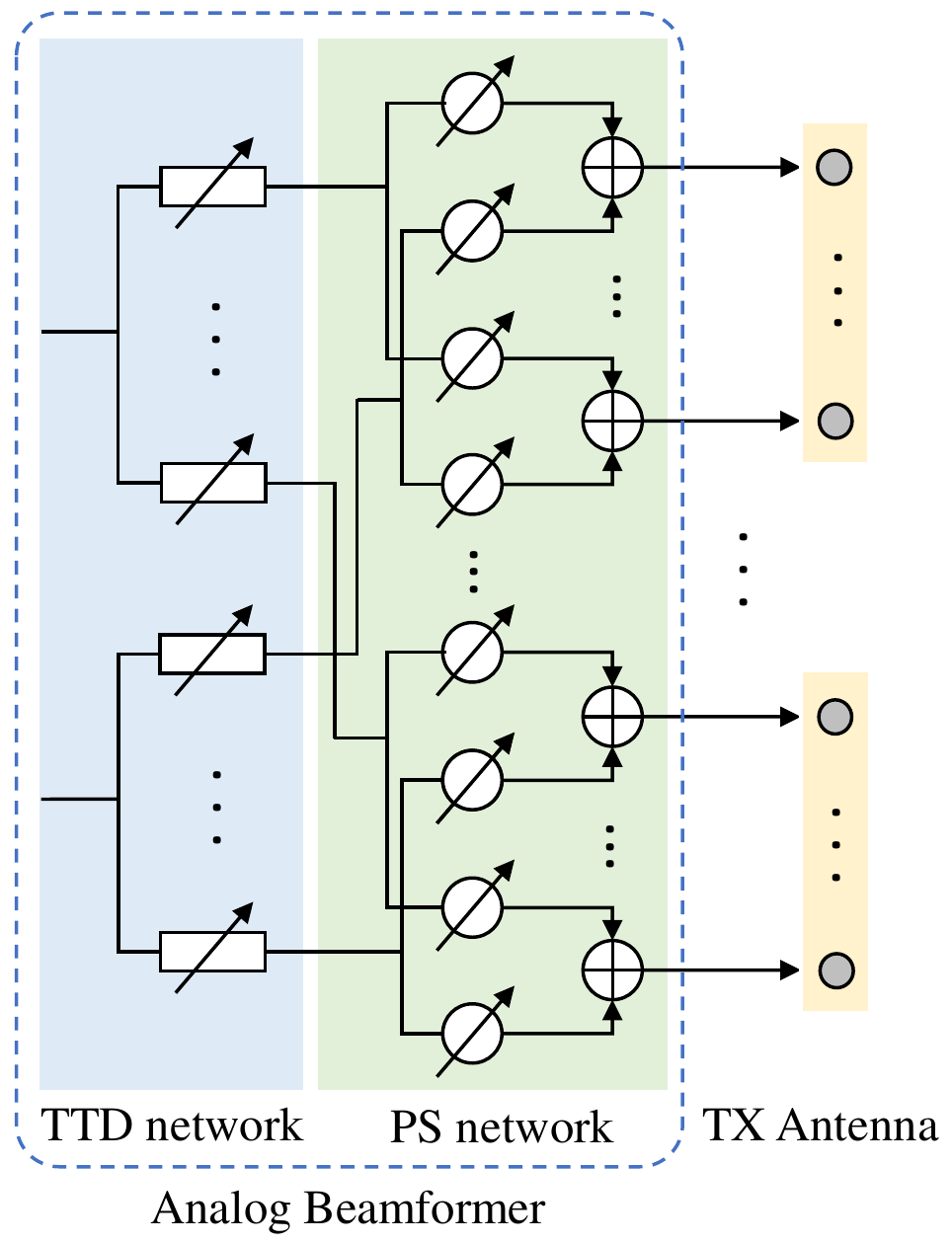}}
	\caption{Illustrations of delay-beam squint in THz-NS-COM and the corresponding true-time-delay network for squint-effect compensation \cite{Tan_jsac_21}, where TDD and PS stand for true time delay and phase shifter, respectively.}
\end{figure*} 

On the other hand, although the atmospheric sparsity leads to marginal molecular absorption effects under inter-NSPS and NSPS-satellite scenarios, it cannot be overlooked for NSPS-airplane data links nearby the bottom of stratosphere, where the density of water vapor is relatively higher than the upper region of the near space. Compared with the terrestrial counterpart, the molecular absorption effects present stronger frequency- and distance- dependency in NSPS-airplane links \cite{Han_cm_18}, where ultra-broadband data transfer with long range over tens of km is expected. This induces additional frequency-selectivity to the propagation channel between NSPS and aerial platforms. Also note that, the traditional terrestrial modeling for molecular absorption loss based on homogeneous assumption of the transmission medium, is no longer valid for THz-NS-COM due to inhomogenity of the high-altitude atmosphere, as stated above \cite{Kokkoniemi_tvt_21}. Hence, mathematical modeling of molecular absorption loss needs substantial modification, which should be a function of transmission distance, center frequency, and the atmospheric temperature and pressure curves across different altitudes. Aside from signal attenuation, the molecular absorption phenomenon also generates colored noise components, termed as molecular absorption noise, which originates from black-body radiation of the atmosphere. The characteristics of such ambient noise depend on the resonant frequencies of different types of atmospheric molecules, which can be modeled by the antenna brightness temperature following the Rayleigh-Jeans law or Planck's law \cite{Kokkoniemi_tvt_21}. 


{\bf Channel Modeling for THz-NS-COM}: Traditional channel modeling strategies for THz communications can be classified into deterministic and stochastic methods \cite{Han_cm_18}. On one hand, the deterministic strategies are dependent on the knowledge of properties of the transmission medium, geometrical information of the propagation environment, the positions as well as relative velocities of the transceiver nodes, etc. Despite its superior accuracy on channel modeling, it is challenging to obtain the aforementioned knowledge especially for the complicated propagation environment of THz-NS-COM. Besides, the deterministic methods usually suffer from high computational overhead, which may be impractical for NSPS-mounted payloads with limited size, weight and power consumption. On the other hand, the stochastic methods only utilizes the empirical channel measurements for channel modeling, which are featured by the geometry-based stochastic channel model (GBSM) widely employed for standard propagation modeling, e.g., the 3rd Generation Partnership Project (3GPP) technical reports. Since detailed knowledge of the propagation environments are no longer required for stochastic methods, the computational complexity can be eliminated, which, however, induces additional channel modeling errors. Artificial intelligence has been recently introduced to physical-layer design such as demodulation and channel estimation, which approaches the performance of classical receivers with lower complexity. Inspired by this concept, the artificial neural network is also expected to support channel modeling of THz-NS-COM in the near future. For instance, the generative adversarial network (GAN) architecture can be invoked to enlarge the training set of channel measurement data. The massive amount of training data is further exploited by a convolutional neural network for channel modeling, which can implicitly characterize a plethora of factors in the propagation environment of THz-NS-COM, such as altitude/distance-dependence and frequency-selectivity of the path loss, the mobility of transceiver terminals, and resultant antenna misalignment fading.

\section{Robust Beamforming for UM-MIMO}
\subsection{Challenges and Key Techniques}
THz-NS-COM is required to support long-range data links over 100 km, which necessitates highly directional transmission with sufficiently large antenna gain. Although the lumbersome parabolic antennas are inapplicable for NSPS with limited payload budget, utilization of the THz frequencies could readily enable ultra-massive antenna arrays to achieve superior beamforming gain with compact size. Nevertheless, the ultra-massive MIMO (UM-MIMO) system for THz-NS-COM suffers from inevitable delay-beam squint effects \cite{Liao_jsac_21}, as illustrated in Fig. \ref{fig3a}. Specifically, when the direction of incoming signals is non-perpendicular to the antenna array, different propagation delays of the received signals can be witnessed at various receive (RX) antennas. Due to extremely high-rate sampling for ultra-broadband THz transmission, the resultant delay gap between antennas could approach several symbol intervals, yielding non-negligible inter-symbol interference (ISI). Such phenomenon is referred to as the delay squint effect. Furthermore, the phase shifters for classical analog beamforming are designed based on the center frequency of the consumed bandwidth, which leads to severe beam split effects (beam squint effects) for THz UM-MIMO with ultra-broad bandwidth. This is because the frequency-independent analog precoder is not aligned with most subcarriers within the employed spectrum, resulting in severe degradation of beamforming gain. To address these technical challenges, a true-time delay unit module was implemented before the analog beamformer against the delay-beam squint effects \cite{Liao_jsac_21}, as presented in Fig. \ref{fig3b}. Explicitly, by inducing specially designed propagation delays to different transmit (TX) antennas, the true-time delay units are capable of counteracting the delay squint effect, and realizing a frequency-dependent phase-shifter network to ensure desirable beam directions across the overall communication bandwidth.

Aside from the delay-beam squint effects, the antenna misalignment issue cannot be overlooked in THz-NS-COM. Due to the use of highly directional beams over ultra-long-range transmission, even slight position/orientation change of the NSPS terminals can result in severe beam misalignment fading, not to mention the civil aircraft or satellite nodes. Besides, unlike the aerial/satellite communication networks, where the flight trajectories of transceiving platforms usually follow predetermined routes, the mobility of NSPS nodes (especially for balloons) is irregular due to elusive wind effects. This puts even larger difficulties on beam tracking for THz-NS-COM than its aerial/satellite counterparts. To address this issue, the beam split effects can be exploited for fast beam training. Explicitly, the coverage of split beams can be flexibly adjusted by true-time delay units to include the spatial directions of the NSPS nodes, which allows beam tracking of multiple spatial directions of transceiver terminals in each time slot, thus accelerating the beam training process. For more details, readers can refer to \cite{Tan_jsac_21} and the references therein.

\subsection{Open Issues}
{\bf Beam Tracking under Scattering and Refraction Effects}: In comparison with lower frequencies, THz signal transmission with ultra-short wavelength suffers from more severe scattering and refraction effects \cite{Chen_cm_21}, which are originated from the existence of aerosol particles and the inhomogeneous atmospheric medium. Despite the sparsity of atmosphere in the near-space environment, the scattering and refraction effects can be significantly accumulated along the ultra-long transmission distance approaching 100 km, which cause non-negligible deviations of the beam directions. Therefore, in the beam tracking process, angular modifications of the transceiver beams are required to counteract the bending effects, based on the properties of the propagation environment, such as the temperature, pressure and atmospheric constituents at different altitudes and latitudes.

{\bf RIS-based UM-MIMO for THz-NS-COM}: Despite the attractive merits, Existing THz-band UM-MIMO systems require considerable radio-frequency (RF) chains and a complex feeding network for analog beamforming. These result in high hardware overhead and energy consumption, which contradict with the limited budget for the NSPS-mounted communication payloads. To this end, the reconfigurable intelligent surface (RIS), a passive meta-surface composed of numerous programmable metallic elements, can be employed for replacement of the classical phased-array antennas \cite{Tang_jsac_20}. By emitting THz electromagnetic waves onto the RIS through irradiation and adjusting the reflecting coefficients of RIS elements, an RF-chain-free data transmission with analog beamforming can be realized without the need of feeding network. Nevertheless, such RIS-assisted transmitter suffers from limited amplitude adjustment of the reflecting coefficients of RIS elements. This adds great difficulties in realizing quadrature amplitude modulation (QAM), causing performance loss at high spectral efficiency. Furthermore, the discrete rotated phases of RIS elements in practical degrades the analog beamforming inevitably, which demands careful investigation to meet the stringent requirement of beam alignment in THz-NS-COM. 

\section{Transceiver Design under Peculiarities of THz-NS-COM}
\subsection{Challenges and Key Techniques} 
\begin{figure}[t!]
	\begin{center}
		\includegraphics[width=1\linewidth, keepaspectratio]{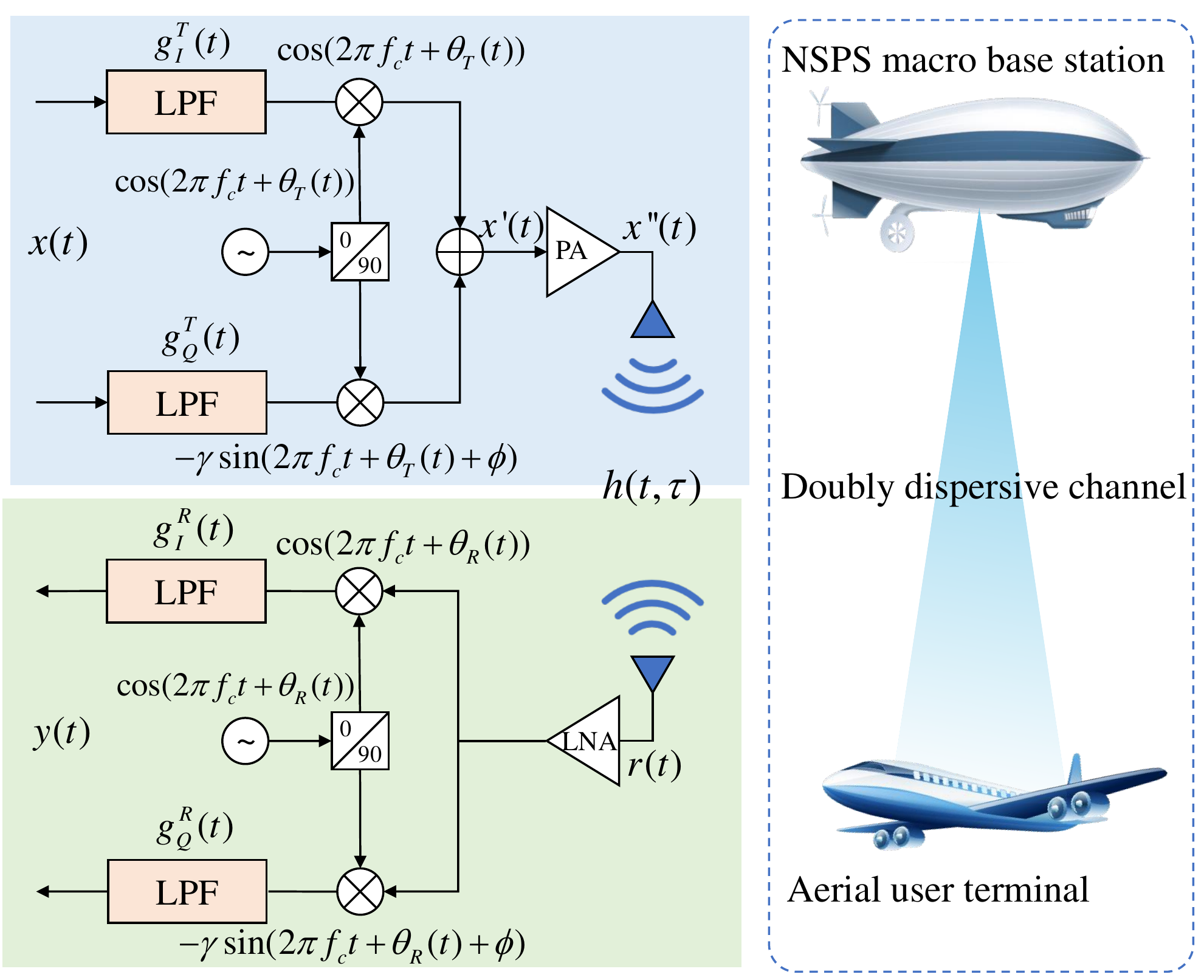}
	\end{center}
	\caption{Diagram of THz-NS-COM transceivers with hardware imperfections, where LPF, PA and LNA denote low-pass filter, power amplifier and low-noise amplifier, respectively. $g_{I(Q)}^{T(R)}(t)$ stand for the impulse responses of the low-pass filters; $\gamma$ and $\phi$ denote amplitude and phase I/Q imbalance coefficients, respectively; $\theta_{T(R)}(t)$ represents phase noise at the transceiver; $h(t,\tau)$ is defined as the time-frequency doubly-dispersive channel.}
	\label{fig4}
\end{figure}

Despite its capability of attaining ultra-broadband data transmission, the THz spectrum has been often recognized as the ``THz Gap'' due to the immature transceiving technology between classical microwave and infrared frequencies. This causes severe hardware imperfections that present frequency-selective wideband characteristics, including phase noise, in-phase/quadrature (I/Q) imbalance, nonlinearity of the power amplifier and frequency multiplier, etc., as illustrated in Fig. \ref{fig4}. Furthermore, the NSPS-mounted transceivers tend to employ the direct-conversion architecture to reduce the implementation cost, which, however, is especially sensitive to the hardware impairment issue. To tackle this problem, existing parameter estimation and compensation techniques for terrestrial THz communications can be also applied to the NSPS-mounted transceiver. Explicitly, at the transmitter, the TX I/Q imbalance and nonlinearity of the power amplifier are required to be jointly estimated before pre-compensation, which is a non-convex problem due to highly coupling of the hardware imperfection parameters. To circumvent this obstacle, these parameters can be estimated via sophisticated alternating iterative methods. At the receiver, a multi-stage post-processing algorithm \cite{Sha_jsac_21} can be invoked to eliminate the remaining RX I/Q imbalance and phase noise from the transceiver: a) Estimating the equivalent channel of physical channel and I/Q imbalance via extended Kalman filtering; b) Decoupling the I/Q imbalance and physical channel by solving a blind deconvolution problem; c) Equalize/compensate the I/Q imbalance, physical channel and phase noise subsequently.

Besides, the propagation channel of THz-NS-COM suffers from non-negligible frequency selectivity and strong Doppler shift effects resulted from extremely high operating frequency and drastic mobility of the user terminals, e.g., civil aircraft, fighter planes and satellites. Hence, a complex time-frequency doubly-dispersive channel is constituted, which becomes even more complicated by incorporating the frequency-selective property of Doppler shifts due to huge communication bandwidth, referred to as Doppler squint effects. These factors pose additional difficulties to the aforementioned channel estimation, decoupling and equalization phases. The novel orthogonal time frequency space (OTFS) philosophy may help to address this issue. To be more specific, the data symbols are modulated in the delay-Doppler domain to obtain better resilience against Doppler shifts. Moreover, additional channel sparsity can be attained by converting the time-frequency channel to delay-Doppler domain, which further simplifies the operations from a) to c).

\subsection{Open Issues} 

{\bf Low-Complexity Transceiver Design}: Existing compensation schemes for wideband hardware imperfections suffer from extremely high computational complexity. This motivates low-complexity transceiver design for THz-NS-COM to alleviate the burdens of NSPS-mounted communication payloads. In the existing literature, the narrow-band hardware imperfections are approximately modeled as additive noise components for simplicity based on the Bussgang Theorem, which requires further investigations by involving the frequency selectivity of wideband I/Q imbalance and phase noise. Besides, the artificial-intelligence-based approaches may also ease the parameter estimation and compensation processes of complex hardware imperfections.

{\bf Optoelectronic Transceivers for THz-NS-COM}: In classical electronic THz communication systems, frequency multipliers are required for generation of the THz-band carriers from lower frequencies, which induce additional nonlinearity from the multiplier chain. Alternatively, We can circumvent the use of nonlinear frequency multipliers by invoking optoelectronic methods for THz-frequency generation, based on direct modulation scheme using quantum cascade lasers (QCL) or heterodyne scheme using a uni-traveling carrier photodiode (UTC-PD) photomixer \cite{Yu_mag_20}. Furthermore, the optoelectronic THz communication system suffers from marginal phase noise due to extremely narrow linewidth of the laser signals. Nevertheless, current optoelectronic methods mainly rely on bulky front-end devices, which are still unsuitable for NSPS applications with limited communication payload carrying capability, thus necessitating compact chip-based design of the optoelectronic THz transceiver.

\section{Integrated Design of Sensing and Communications} 
\subsection{Concepts and Key Techniques}
Aside from broadband data transmission, the THz signal is also capable of target sensing with superior resolution of ranging and velocity estimation, attributed to its ultra-wide bandwidth and THz-scale carrier frequency, respectively. Since the path loss is substantially weaker in the near-space propagation environment than its terrestrial counterpart, THz frequencies could realize accurate radar sensing over 100 km. Such detection range can be further extended by the use of UM-MIMO with highly directional beamforming, which also supports simultaneous sensing at various directions by generating multiple TX beams. Therefore, THz radar signals can enable numerous military/civilian applications of NSPS, e.g., reconnaissance, navigation and positioning. To satisfy the constraint on the size of NSPS-mounted payloads, the sensing and communication subsystems can be fused together by sharing the same set of hardware platform, which usually require an integrated waveform design to support high-rate data transmission and accurate radar sensing concurrently. Despite the randomness of the communication signals, orthogonal frequency division multiplexing (OFDM) can attain perfect auto-correlation property by employing constant-amplitude modulation formats, making it suitable for radar sensing. However, the implementation of OFDM inevitably induces complex transceiver structure and extra power backoff of the power amplifier due to high peak-to-average power ratio (PAPR), which is especially undesirable for NSPS-mounted THz front ends. To tackle this issue, a multi-subband waveform for joint radar and communications is developed based on the philosophy of distributed single-carrier frequency division multiple access (SC-FDMA) \cite{Mao_tcom_22}. Explicitly, the overall spectrum is divided with multiple non-overlapped subbands, where a radar sequence with perfect correlation property and the data symbol components are modulated at non-intersect frequency points. By such arrangement, the ISAC waveform can inherit good correlation property from the radar component, whilst at the same time can ensure broadband communications without mutual interference. Besides, only several narrow-band analog-to-digital converters (A/Ds) are required for sampling without the need of full-band A/D, which poses great difficulties for fabrication under ultra-broad bandwidth. 

Despite the low mobility of NSPS, the THz-scale center frequency and large velocities of the aerial user terminals contribute to pronounced Doppler shift, which also degrades the sensing performance besides data transmission. To be more specific, additional dominant range sidelobes can be induced by strong Doppler shift effects, leading to higher false alarm rate and reduced peak-to-sidelobe level of the radar range profile. To address this issue, differential encoding can be invoked to counteract the Doppler shift effects by multiplications between a received sample and the conjugate of its neighboring one. However, such method becomes invalid at low signal-to-noise ratios (SNRs) attributed to amplification of the noise components. Alternatively, the waveform parameters are specially designed to adjust the positions of dominant range sidelobes based on the number theory, which eliminates possible false alarms by moving these sidelobes outside the region of interest \cite{Mao_tcom_22}.

\subsection{Open Issues}
{\bf Collaborations of Sensing and Communications}: Instead of treating each other as interferers, the performance of THz-NS-COM can be enhanced by highly collaborations of both functions. Specifically, THz-NS-COM systems are especially prone to beam misalignment fading due to the ``razor sharp'' beam shape and unpredictable mobility of NSPS terminals. To alleviate this issue, the THz radar detection results with ultra-high resolution can be utilized to assist existing beam alignment strategies for higher accuracy and time efficiency, which requires detailed researches in the future.

\begin{figure}[t!]
	\begin{center}
		\includegraphics[width=1\linewidth, keepaspectratio]{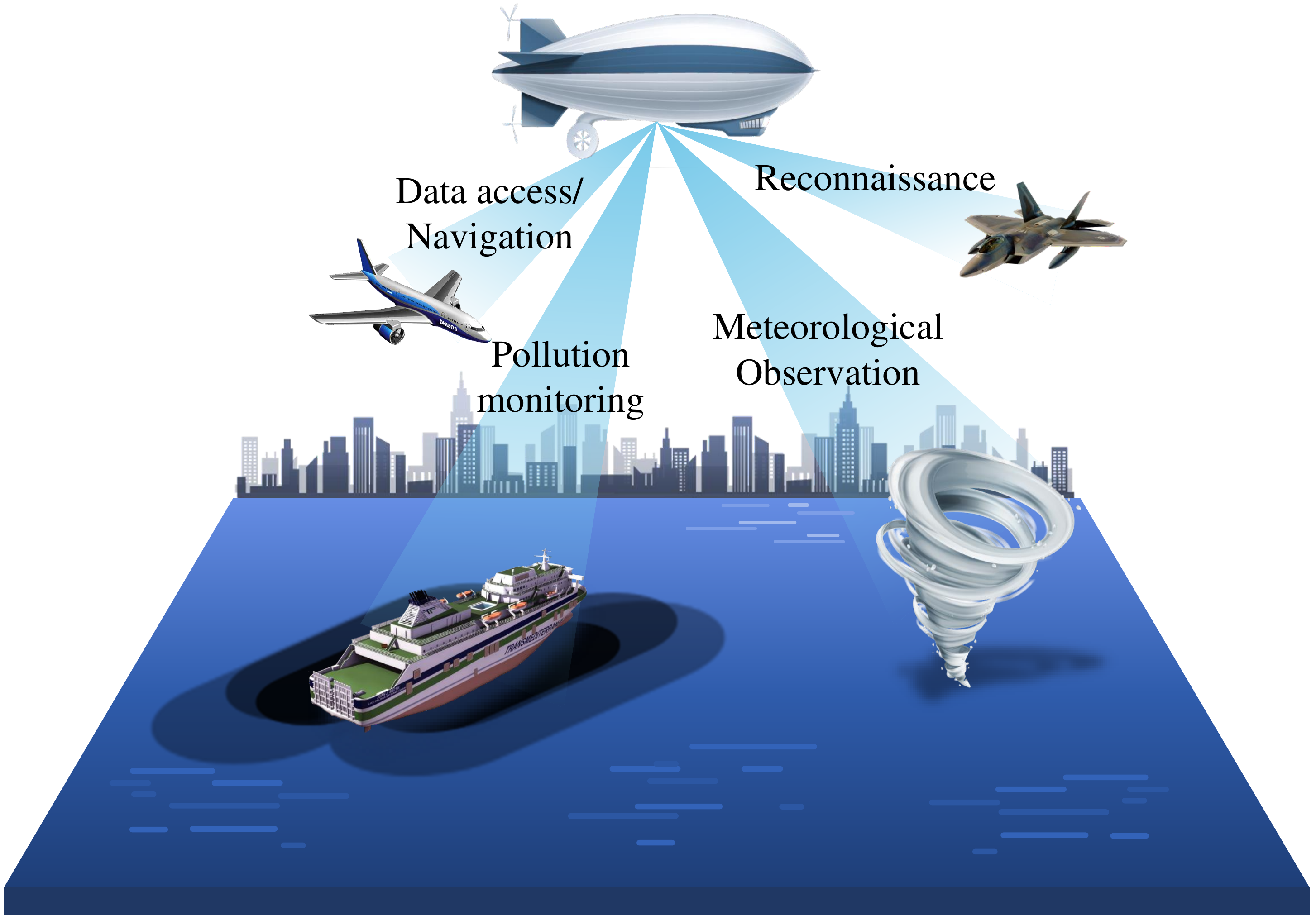}
	\end{center}
	\caption{Different radar applications of NSPS, where THz imaging or spectroscopy technologies can be employed for superior sensing performances.}
	\label{fig5}
\end{figure}

{\bf Electromagnetic Compatibility (EMC) Issue}: As shown in Fig. \ref{fig5}, numerous sensing-related missions can be realized by additional NSPS-mounted radars/sensors aside from the ISAC payload, such as pollution monitoring, scientific observation and military reconnaissance. The aforementioned tasks rely on different pivotal technologies like synthetic aperture radar (SAR) imaging and molecular line spectroscopy, which can attain superior performances when operating at THz frequencies. By such arrangement, it will be essential to consider the EMC issue between ISAC devices of THz-NS-COM and the other THz-band instruments, in case of possible overlap of their used frequency bands.


\section{Conclusion}
In this paper, the novel THz-NS-COM philosophy was investigated from the physical-layer perspective. Explicitly, the prospects of THz-NS-COM were envisioned by highlighting its attractive merits such as capabilities of realizing ultra-broadband long-range communications and superior ISAC performance, which, however, still faces significant challenges for its practical implementation. To provide some design guidelines for THz-NS-COM, the characteristics of the propagation channel and its modeling methods were firstly discussed. Afterwards, the technical barriers and potential technologies of THz-NS-COM were specified from three aspects of beamforming for UM-MIMO, transceiver signal processing against distortions and ISAC applications, respectively. Meanwhile, the corresponding open issues were also provided for future investigation.


\begin{thebibliography}{1}
\bibitem{Zhang_conf_07}
Z. Bo, R. Qinghua, L. Yunjiang, C. Zhenyong, and Z. Feng, ``Characteristic and simulation of the near space communication channel,'' \emph{Proc. International Symposium on Microwave, Antenna, Propagation and EMC Technologies for Wireless Communications}, Hangzhou, China, Aug. 2007, pp. 769-773. 

\bibitem{StationX_2022} 
\emph{X-Station}. Accessed: Mar. 10, 2022. [Online]. Available: \url{http://www.stratxx.com/xstation.html}

\bibitem{Loon_Google_2022} 
\emph{Loon: Expanding internet connectivity with stratospheric balloons}. Accessed: Mar. 10, 2022. [Online]. Available: \url{https://x.company/projects/loon/}

\bibitem{Kurt_survey_21}
G. Karabulut Kurt \emph{et al.}, ``A vision and framework for the high altitude platform station (HAPS) networks of the future,'' \emph{IEEE Commun. Surveys \& Tut.}, vol. 23, no. 2, pp. 729-779, Secondquarter 2021.

\bibitem{Chen_cm_21}
Z. Chen, C. Han, Y. Wu, L. Li, C. Huang, Z. Zhang, G. Wang, and W. Tong, ``Terahertz wireless communications for 2030 and beyond: A cutting-edge frontier,'' \emph{IEEE Commun. Mag.}, vol. 59, no. 11, pp. 66-72, Nov. 2021.

\bibitem{Mao_tcom_22}
T. Mao, J. Chen, Q. Wang, C. Han, Z. Wang, and G. K. Karagiannidis, ``Waveform design for joint sensing and communications in millimeter-wave and low terahertz bands,'' \emph{IEEE Trans. Commun.}, to appear.


\bibitem{Saeed_jsac_21}
A. Saeed, O. Gurbuz, A. O. Bicen, and M. A. Akkas, ``Variable-bandwidth model and capacity analysis for aerial communications in the terahertz band,'' \emph{IEEE J. Sel. Areas Commun.}, vol. 39, no. 6, pp. 1768-1784, Jun. 2021.

\bibitem{Liao_jsac_21}
A. Liao, Z. Gao, D. Wang, H. Wang, H. Yin, D. W. K. Ng, and M. Alouini, ``Terahertz ultra-massive MIMO-based aeronautical communications in space-air-ground integrated networks,'' \emph{IEEE J. Sel. Areas Commun.}, vol. 39, no. 6, pp. 1741-1767, Jun. 2021.	

\bibitem{britannica_2022} 
\emph{Layers of Earth's atmosphere}. Accessed: Mar. 24, 2022. [Online]. Available: \url{https://www.britannica.com/science/atmosphere/Troposphere#/media/1/41364/99826}

\bibitem{Han_cm_18}
C. Han and Y. Chen, ``Propagation modeling for wireless communications in the terahertz band,'' \emph{IEEE Commun. Mag.}, vol. 56, no. 6, pp. 96-101, Jun. 2018.

\bibitem{Kokkoniemi_tvt_21}
J. Kokkoniemi, J. M. Jornet, V. Petrov, Y. Koucheryavy, and M. Juntti, ``Channel modeling and performance analysis of airplane-satellite terahertz band communications,'' \emph{IEEE Trans. Veh. Technol.}, vol. 70, no. 3, pp. 2047-2061, Mar. 2021.

\bibitem{Tan_jsac_21}
J. Tan and L. Dai, ``Wideband beam tracking in THz massive MIMO systems,'' \emph{IEEE J. Sel. Areas Commun.}, vol. 39, no. 6, pp. 1693-1710, Jun. 2021. 

\bibitem{Tang_jsac_20}
W. Tang \emph{et al.}, ``MIMO transmission through reconfigurable intelligent surface: System design, analysis, and implementation,'' \emph{IEEE J. Sel. Areas Commun.}, vol. 38, no. 11, pp. 2683-2699, Nov. 2020.

\bibitem{Sha_jsac_21}
Z. Sha and Z. Wang, ``Channel estimation and equalization for terahertz receiver with RF impairments,'' \emph{IEEE J. Sel. Areas Commun.}, vol. 39, no. 6, pp. 1621-1635, Jun. 2021.

\bibitem{Yu_mag_20}
L. Zhang, X. Pang, S. Jia, S. Wang, and X. Yu, ``Beyond 100 Gb/s optoelectronic terahertz communications: Key technologies and directions,'' \emph{IEEE Commun. Mag.}, vol. 58, no. 11, pp. 34-40, Nov. 2020.


\end{thebibliography}
\end{document}